\documentclass[aip,reprint]{revtex4-1}
\usepackage{graphicx}
\usepackage{dcolumn}
\usepackage{bm}
\usepackage{mathtools}
\usepackage{amsmath}
\usepackage{color}
\usepackage{natbib}
\usepackage[colorlinks,plainpages=false,linkcolor=black,urlcolor=blue,citecolor=black,pdfpagemode=UseNone,pdfstartview=FitBH]{hyperref}
\usepackage{color,soul}
\usepackage[framemethod=tikz]{mdframed}

\begin{document}

\title{Comparative study of LaNiO$_{\text{3}}$/LaAlO$_{\text{3}}$ heterostructures grown by pulsed laser deposition and oxide molecular beam epitaxy}

\author{F. Wrobel}
\affiliation{Max Planck Institute for Solid State Research, Heisenbergstr. 1, 70569 Stuttgart, Germany}

\author{A. F. Mark}
\affiliation{Max Planck Institute for Solid State Research, Heisenbergstr. 1, 70569 Stuttgart, Germany}

\author{G. Christiani}
\affiliation{Max Planck Institute for Solid State Research, Heisenbergstr. 1, 70569 Stuttgart, Germany}

\author{W. Sigle}
\affiliation{Max Planck Institute for Solid State Research, Heisenbergstr. 1, 70569 Stuttgart, Germany}

\author{H.-U. Habermeier}
\affiliation{Max Planck Institute for Solid State Research, Heisenbergstr. 1, 70569 Stuttgart, Germany}

\author{P. A. van Aken}
\affiliation{Max Planck Institute for Solid State Research, Heisenbergstr. 1, 70569 Stuttgart, Germany}

\author{G. Logvenov}
\affiliation{Max Planck Institute for Solid State Research, Heisenbergstr. 1, 70569 Stuttgart, Germany}

\author{B. Keimer}
\affiliation{Max Planck Institute for Solid State Research, Heisenbergstr. 1, 70569 Stuttgart, Germany}

\author{E. Benckiser}
\email[]{E.Benckiser@fkf.mpg.de}
\affiliation{Max Planck Institute for Solid State Research, Heisenbergstr. 1, 70569 Stuttgart, Germany}

\date{\today}

\begin{abstract}
Variations in growth conditions associated with different deposition techniques can greatly affect the phase stability and defect structure of complex oxide heterostructures. We synthesized superlattices of the paramagnetic metal LaNiO$_3$ and the large band gap insulator LaAlO$_3$ by atomic layer-by-layer molecular beam epitaxy (MBE) and pulsed laser deposition (PLD) and compared their crystallinity, microstructure as revealed by high-resolution transmission electron microscopy images and resistivity. The MBE samples show a higher density of stacking faults, but smoother interfaces and generally higher electrical conductivity. Our study identifies the opportunities and challenges of MBE and PLD growth and serves as a general guide for the choice of deposition technique for perovskite oxides.
\end{abstract}

\maketitle

The prospect of realizing new functional devices based on transition metal oxides relies on the possibility to tune their electronic or magnetic properties by epitaxial strain, confinement, doping, or interface effects. For precise control of these parameters, a high crystal quality, accurate stoichiometry, and atomically sharp interfaces are required. In the past years, thin film deposition techniques such as PLD, magnetron sputtering or MBE have been optimized for the growth of oxide heterostructures.~\cite{Boris2011,Scherwitzl2009,Bozovic2002,King2014, Hauser2015} The growth conditions differ substantially in deposition temperature, pressure, energy of the impinging particles and stoichiometry control. So far little is known about how imperfections, such as interface roughness, chemical non-stoichiometry, surface degradation, and the intergrowth of different phases are connected to different growth methods and how they influence the physical properties of the resulting thin-film structures.

Here we report on a comparative investigation of an intensively studied model system, namely superlattices (SLs) composed of perovskite-type LaNiO$_3$ (LNO) and LaAlO$_3$ (LAO), a paramagnetic metal and a large band gap insulator in bulk, respectively.~\cite{Boris2011,Benckiser2011,Freeland2011,Detemple2011,Frano2013,Wu2013,Kinyanjui2014,Lu2016}
Prior work has revealed a transition from a bulk-like paramagnetic state to a spin-density-wave phase with reduced electrical conductivity with decreasing thickness of the LNO layers.~\cite{Boris2011,Frano2013} While the properties of both sets of SLs are generally in good agreement, our systematic study revealed quantitative differences in the value of resistivity and in the microstructure which can be attributed to lower interfacial roughness in MBE samples and a more accurate lanthanum-to-nickel ratio in PLD samples.

\begin{figure}[tb]
\center\includegraphics[width=0.9\columnwidth]{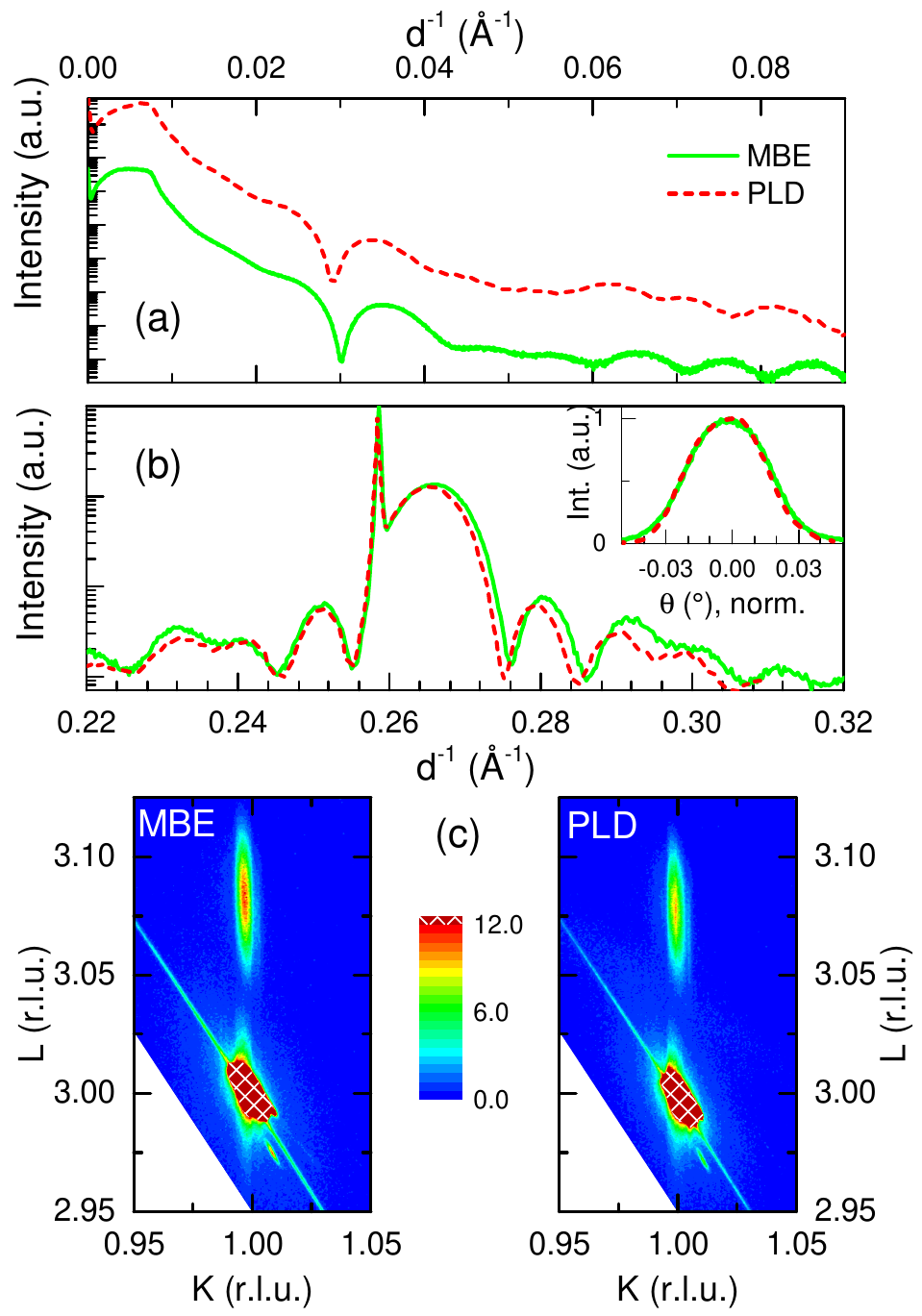}
\caption{(Color online) Hard x-ray data of two nominally identical $[4/4]\times 3$ samples on LSAT substrates grown by MBE (green, solid line) and PLD (red, dashed line). (a) Reflectivity. (b) Scans through the (001) reflections of substrate and SL. The average $c$-axis parameter is 3.768(8)\,\r{A} for the MBE sample and 3.773(5)~\r{A} for the PLD sample. The inset shows the corresponding rocking curves measured at the (001) reflection of the film where the intensities were normalized such that the maximum intensity of each scan is 1. (c) Reciprocal space maps around the (103) reflections of substrate and SL show that both samples are fully strained to the substrate.}\label{XRD}
\end{figure}

We compare SLs with the composition [(LNO)$_{{n}}$/(LAO)$_{{m}}]_{{l}}$, abbreviated by $[n/m]\times l$ in the following. They were deposited on the (001) surfaces of cubic (LaAlO$_{\text{3}}$)$_{\text{0.3}}$-(Sr$_{\text{2}}$AlTaO$_{\text{6}}$)$_{\text{0.7}}$ (LSAT), with a small lattice mismatch, and on the (001) surfaces of SrTiO$_3$ (STO), with larger mismatch. The bulk (pseudo)cubic lattice constants of STO, LSAT, LNO and LAO are 3.91, 3.87, 3.84 and 3.83 \r{A}, respectively.~\cite{Wold1957} Growth parameters were optimized individually for each growth technique. PLD samples were prepared as described in Ref.~\onlinecite{Wu2013}. MBE samples were grown at $550^{\circ}{\rm C}$ in a 2.5$\times$10$^{\text{-5}}$\,mbar ozone atmosphere and cooled down at the same pressure. The fluxes of the effusion cells were calibrated with a quartz crystal monitor before growth. Reflection high-energy electron diffraction was used for real-time monitoring, enabling atomic layer control. A detailed description of the MBE system and the growth process can be found in the Supplementary Material and in Ref.~\onlinecite{Baiutti2014}.

High-resolution XRD and reflectivity measurements were performed using a four-circle diffractometer with a Cu $K_{\alpha 1}$ source. Transport measurements (sheet resistance and Hall resistance) were conducted in van-der-Pauw geometry in fields up to 9\,T. Where needed, the Hall resistance $\rho$$_\text{H}$ was corrected for the magnetoresistance, i.e., \(\rho_\text H = (\rho_{xy} ({+B}) - \rho_{xy}({-B}))/2\). The Hall coefficient was calculated by \(\text R_\text H = t \cdot \partial \rho_\text H / \partial \text B \) where the thickness \textit{t} is the LNO \textit{c}-axis parameter multiplied by the number of LNO layers.

\begin{figure}[tb]
\center\includegraphics[width=1\columnwidth]{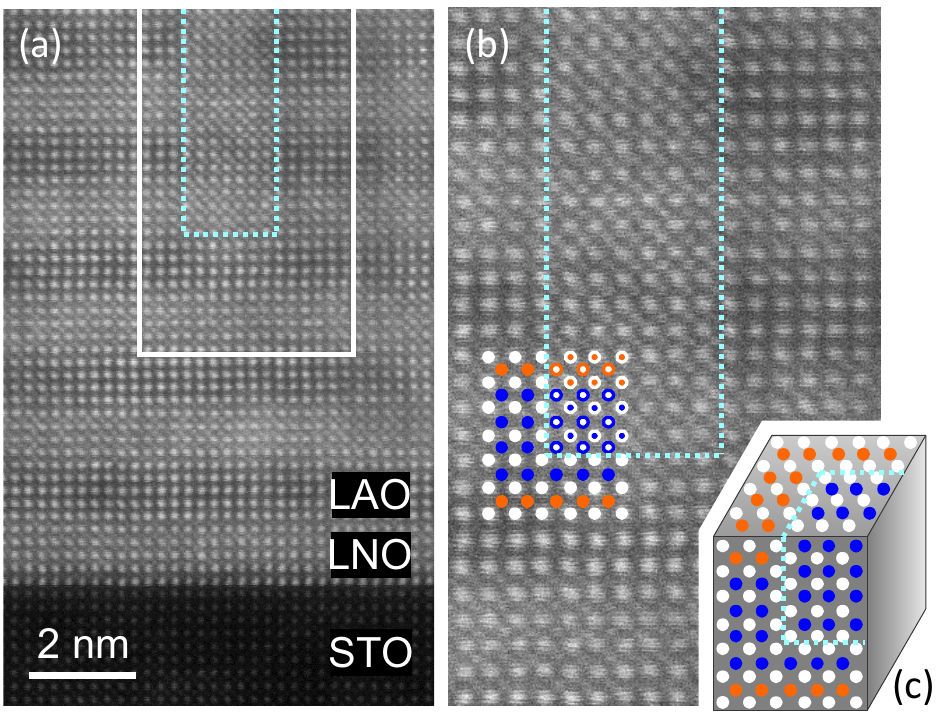}
\caption{(Color online) Annular dark-field (ADF) image of an MBE-grown $[4/4]\times8$ SL on STO where the dashed blue lines mark the edges of a 3D-Ruddlesden-Popper (3D-RP) fault. (a) Full image and (b) enlarged view of the boxed area in (a). (c) A 3-dimensional representation of a 3D-RP fault. The 3D-RP fault arises from an additional LaO layer at its bottom such that all layers above are shifted by half a unit cell. In a TEM image, a projection of several atoms on top of each other is measured and thus, the faulted area appears as an overlay of the ideal structure -- represented by big dots in (b) -- and the shifted structure -- represented by small dots in (b). White, blue and orange dots represent La, Ni, and Al, respectively. Note that the 3D-RP faults start in the LNO layers.}\label{TEM_STO}
\end{figure}

Transmission electron microscopy (TEM) was done using scanning mode and the annular dark field (ADF) detector in an abberation corrected JEOL ARM200F operated at 200 kV. Each thin foil was prepared by first cutting a cross section through the layers, then thinning the section using mechanical grinding and polishing on both sides to create a wedge-shaped slice (wedge angle 2$\,^{\circ}$). The slice was thinned to electron transparency by ion-milling with LN$_{\text{2}}$ cooling.

Thin SLs with the composition $[4/4]\times 3$ grown by MBE and PLD on LSAT show only minor differences in their XRD patterns, both in scans through the (0 0 1) and (0 1 3) reflection (in pseudo-cubic notation) and in reflectivity measurements (Fig.~\ref{XRD}). The LNO/LAO-averaged pseudo-cubic in- and out-of-plane lattice parameters are identical within the experimental error. From the damping of the reflectivity curves, we also extract nearly identical surface roughnesses of 4\,\AA\ (about one perovskite unit cell) for both samples. Differences between MBE and PLD grown samples become more apparent in x-ray diffraction/reflectivity data of thicker SLs with a larger number of interfaces (e.g. [2/2]x20; see the Supplementary Material).

\begin{figure*}[tb]
\includegraphics[width=\textwidth]{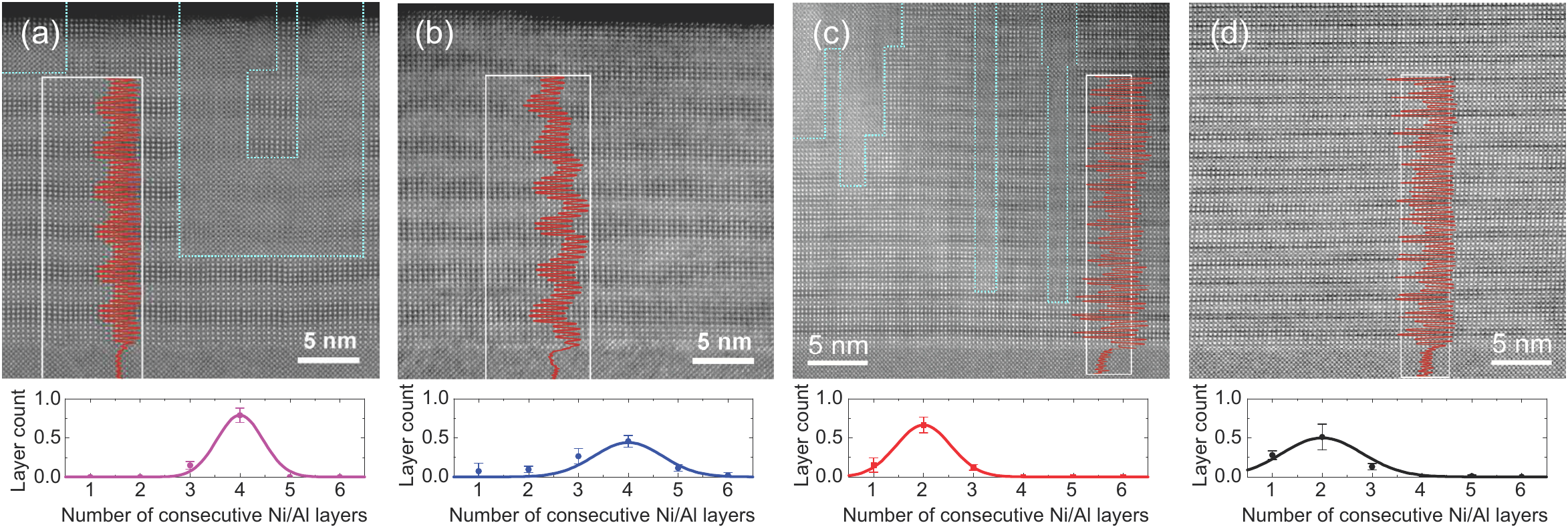}
\caption{(Color online) Annular dark field (ADF) images of four SLs grown on LSAT. The edges of the 3D-RP faults are marked with (blue) dashed lines. (a) $[4/4]\times 8$ (MBE), (b) $[4/4]\times 8$ (PLD), (c) $[2/2]\times 20$ (MBE) and (d) $[2/2]\times 20$ (PLD). The intensity inside the white boxes was integrated in the lateral direction and the resulting profile is shown as red lines inside the boxes. From these plots, we extracted the normalized Ni and Al intensities by dividing each minimum of the plot, which corresponds to one Ni or Al layer, by the maximum of the La layer underneath (labeled by $I_{\rm Ni}$ and $I_{\rm Al}$ in the following). Subsequently, the mean intensity value ($\overline{I}$) is calculated from the average of all normalized Ni and Al intensities along the SL stacking direction. Finally, we counted the number of consecutive Ni ($I_{\rm Ni}>\overline{I}$) and Al ($I_{\rm Al}<\overline{I}$) layers. Layers with an intensity that was equal to the average intensity were not counted. We conducted this kind of analysis for several images of the same sample and the resulting fractions of layers with their corresponding thicknesses are shown in the bottom panels below each ADF image. The solid lines serve as a guide to the eye.}
\label{TEM}
\end{figure*}

\begin{figure*}[tb]
\center\includegraphics[width=1\textwidth]{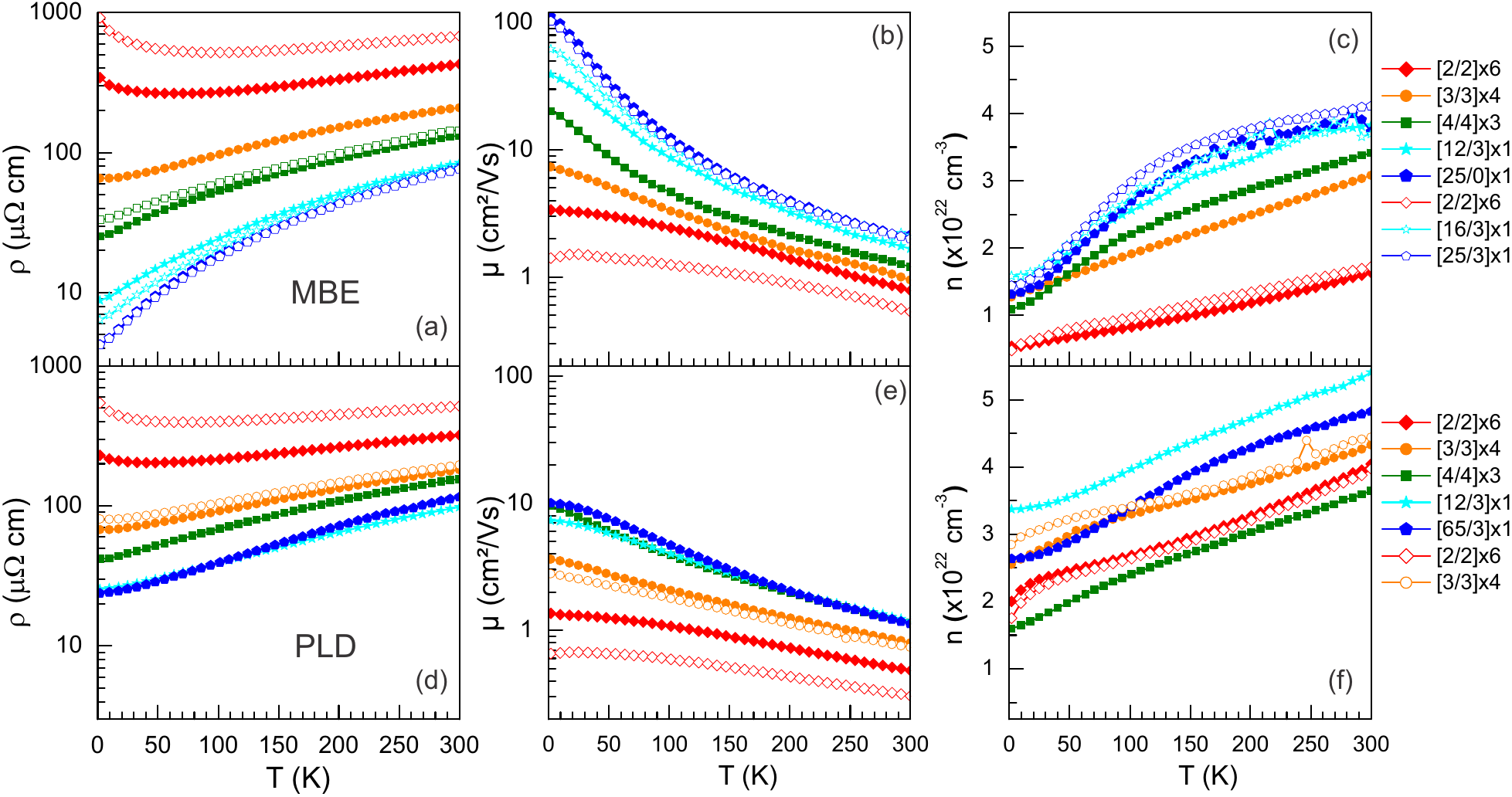}
\caption{(Color online) Temperature dependent resistivity (a \& d), Hall mobility (b \& e), and carrier densities (c \& f) of [(LNO)$_{\textit{n}}$/(LAO)$_{\textit{m}}$]$_{\textit{l}}$ grown by MBE (upper panels, a-c) and PLD (lower panels, d-f) on LSAT with \textit{n}, \textit{m} and \textit{l} as denoted in the legend. Data for nominally identical, PLD- and MBE-grown samples are shown. Data for selected, additional samples are shown with open symbols to demonstrate sample-to-sample variations and systematic differences between PLD and MBE.}
\label{carrier}
\end{figure*}

While the average structure from XRD of the two samples is nearly identical, there are differences at the local level as revealed by TEM. Fig.~\ref{TEM_STO} shows an ADF image taken of an MBE-grown $[4/4]\times 8$ SL on STO. As a first observation we note that the interface with the substrate is atomically sharp. It was previously reported that nanometer-sized NiO precipitates form at the LNO/STO interface due to the difference in polarity.~\cite{Detemple2011} These precipitates were not observed in MBE-grown samples (Fig.~\ref{TEM_STO}). A possible explanation is the difference in deposition sequence. During a laser pulse La, Ni and O are deposited simultaneously on the substrate. The electric potential at the polar interface is not screened such that the formation of NiO is favorable, and NiO precipitates form irreversibly. In MBE growth first a complete La layer is thermally evaporated and oxidized. The resulting LaO layer acts as a buffer that prevents the formation of NiO precipitates. Instead, a smooth layer forms and the polar discontinuity can be alleviated through, e.g., the formation of oxygen vacancies.

Common features observed in ADF images of all samples (Fig.~\ref{TEM_STO} and ~\ref{TEM}) are 3D-Ruddlesden-Popper (3D-RP) faults caused by an additional layer of LaO over a small area-fraction of a layer. \cite{Detemple2012} All subsequent layers are shifted by half a unit cell, and the shifted volume is surrounded by an additional layer of LaO at each side of the fault as can be seen in the enlarged view of Fig.~\ref{TEM_STO}b and in the three-dimensional representation of such a fault in Fig.~\ref{TEM_STO}c. The edges of this 3D-RP are marked with (blue) dashed lines in Fig.~\ref{TEM_STO}a \& b and Fig.~\ref{TEM}a \& c.

Within the shifted volume of a 3D-RP fault, the SL follows the ideal structure. Therefore, in order to estimate the volume of the defect itself, the edges of this shifted volume should be measured. Based on an evaluation of the edges of the faults (marked with (blue) dashed lines in Fig.~\ref{TEM_STO}a \& b and Fig.~\ref{TEM}a \& c), the faulted volume is estimated to be less than 9~$\%$ and 2~$\%$ in the MBE and PLD samples, respectively. These numbers are obtained by measuring the contour lengths of the RP faulted areas in all available images and normalizing them to the total image area.

The LaO layers surrounding the fault in growth direction, i.e., perpendicular to the substrate-film interface, are a consequence of the first additional atomic LaO layer at the bottom of the shifted volume. Therefore, we also compared $R$, defined as the lengths of the faults parallel to the substrate normalized to the total length of the image multiplied by the nominal, total number of LaO layers (64 in the $[4/4]$ and 80 in the $[2/2]$ SLs). In MBE samples, $R$ is less than 1\%. For PLD samples, $R$ is at least a factor of three smaller.

The obtained numbers reflect the La non-stoichiometry necessary to create the observed density of RP faults. Despite the large volume of the shifted areas in the image, the change in stoichiometry indicated by one RP fault is very small because the adjustments in the cation ratio occur only at the edges of the shifted region. In PLD growth the cation stoichiometry is predetermined macroscopically by the composition of the target. In MBE growth stoichiometry control takes place on the level of individual atomic layers and is therefore less accurate, hence a higher density of 3D-RP faults results.

At the substrate-film interface of samples grown on LSAT (Fig.~\ref{TEM}a-d), the images show a bright layer indicating the accumulation of heavier atoms in all samples. In the MBE samples, only one atomic layer right at the interface is affected whereas in the PLD samples 2 -- 3 layers appear brighter. A scenario where statistically distributed vacancies are filled with lanthanum or nickel appears very likely. LSAT is a solid solution with statistical occupation of Sr and La cations at A sites and Al and Ta on B cation sites of the perovskite ABO$_3$ structure, hence the termination of the substrate surface is chemically not well defined. During the MBE growth of the first atomic La layer the vacancies are filled resulting in one bright atomic layer. The extended bright region in the PLD-grown sample can be understood if one takes into account that the energy of impinging particles is much higher, thus more atomic layers are affected.

The lower energy of the impinging particles and the atomic layer-by-layer growth are assumed to also lead to sharper LNO-LAO interfaces, and indeed, the MBE samples appear to have flatter interfaces than the PLD samples. In order to visualize and quantify this observation, we extracted the Ni, Al, and La layer intensities through integrating over the area marked with a white box in Fig.~\ref{TEM}a-d. The statistical analysis of the resulting intensity oscillations in stacking direction are shown in the panels below each ADF image of Fig.~\ref{TEM}a-d. While in the MBE samples about 75\% of the layer stacks have the intended thickness, this is true for only about 50\% in the PLD-grown SLs. The majority of the remaining layer stacks have either one layer less (about 27\% for PLD- and 15\% for MBE-grown SLs) or one layer extra (about 12\% for PLD- and 6\% for MBE-grown SLs). Whereas in PLD-grown samples in average 18\% of the layer stacks deviate by more than one layer in thickness, such strong variations of thickness were not found in MBE-samples. These numbers indicate that the variation of the layer thicknesses is clearly reduced for the MBE samples and that more layers have the intended thickness.

We now compare temperature dependent resistivity, Hall carrier densities and Hall mobility of nominally identical samples grown on LSAT by the two different techniques, which were measured by standard four-point contact experiments (Fig.~\ref{carrier}).

Overall, the resistivity values of PLD samples grown on LSAT is higher and the mobility lower than those of their MBE counterparts (see Figure~\ref{carrier}a, b, d, e). In particular, the low-temperature behavior of the samples with higher LNO thickness deviates significantly which is reflected in a higher residual resistance ratio [$\rho$(300K)/$\rho$(2K)] (RRR) for the MBE samples (e.g. 5.3 vs. 3.7 for a pair of two $[4/4]\times 3$ samples). A 25 u.c. thick film grown by MBE shows an RRR as high as 18.4 and a room temperature resistivity as low as 78~$\mu$Ohm\textperiodcentered cm (Fig.~\ref{carrier}a) which is an improvement compared to previously reported data for both ceramic and thin film samples.~\cite{Zhou2014,King2014} In contrast, the highest RRR found in a PLD sample of the present study is 4.9. These differences also become evident in the Hall mobility of Fig.~\ref{carrier}b \& e and point to higher defect and/ or impurity density in the PLD samples. The presence of 3D-RP faults in the MBE samples has a smaller impact on electrical transport. This can be explained by the fact that the faults rarely start at the sample-substrate interface but mostly after 16 unit cells. Moreover, the size of the shifted volume is larger than the mean free path of charge carriers determined for similar SLs.~\cite{Boris2011} Finally, oxygen vacancies, whose density is hard to determine experimentally in thin films, can lead to a further increase of resistivity.

A dimensional crossover from a paramagnetic metallic state to a weakly insulating, antiferromagnetic state has been observed in LNO-based SLs and thin films on various substrates,~\cite{Scherwitzl2011,Boris2011,Frano2013} and has been attributed to changes at the Fermi level seen in photoemission.~\cite{Yoo2013,King2014,Berner2014} Although the critical thickness  below which these changes occur is under dispute, similar phenomena have been observed in all experiments, on ultra-thin films and SLs alike.

Transport data shown in Fig.~\ref{carrier} confirm the dimensional crossover. In accordance with literature data,~\cite{Boris2011,King2014,Kumah2014} we observe a clear change in resistivity between 2 and 4 unit cell thick LNO layers, while the variations are small for thicker LNO layers. This observation is true for both PLD- and MBE-grown samples, but only in MBE-grown samples the carrier density is strongly reduced (Fig.~\ref{carrier}c) simultaneously. Although LNO possesses a large hole and small electron Fermi surface, it is assumed that the main contribution to conductivity arises from holes, in agreement with field-dependent Hall measurements showing a linear and positive slope.

In MBE samples, the temperature and thickness dependence of the carrier density is highly reproducible and systematically changing with the LNO thickness, as shown with the additional data on similar or nominally identical samples (Fig.~\ref{carrier}c, solid symbols compared with open symbols). This is, in general, not the case for PLD samples (Fig.~\ref{carrier}f). PLD samples do not show a clear trend of the carrier density with respect to the LNO layer thickness. Only the temperature dependence of a thick LNO film resembles its MBE equivalent. Local scale defects such as interface and surface roughness can reduce the effectively conducting thickness and lateral inhomogeneities render the current path more complex. These are possible reasons for the discrepancies between MBE and PLD SLs. For thick films interface and surface roughnesses play a minor role and as a consequence, the carrier density of such a thick film (63 u.c.) agrees with a corresponding MBE sample (25 u.c.).

In conclusion, our comparative study of LNO-LAO SLs grown by MBE and PLD shows the opportunities and challenges of both methods and demonstrates the successful growth of perovskite oxide heterostructures. While the atomic layer-by-layer deposition in MBE yields on average sharper interfaces, the cation stoichiometry is more difficult to control. Depending on the chemistry of the material, the deviations from stoichiometry can give rise to a complex microstructure, as demonstrated by the RP faults occurring in the LNO layers of the MBE film. Future work should focus on mitigating the stoichiometric variations in MBE growth and on transport measurements on samples with reduced lateral dimensions.

\section*{supplementary material}
See Supplementary Material for additional information on the sample growth as well as for further XRR and XRD data.

\begin{acknowledgments}
We acknowledge the financial support by the DFG via TRR80, project G1, and funding from the European Union Seventh Framework Program [FP/2007-2013] under grant agreement no 312483 (ESTEEM2).
\end{acknowledgments}

\end{document}